\documentclass[amsmath,amssymb,pra,aps,twocolumn,mathbbm,superscriptaddress]{revtex4}
\usepackage{amssymb}
\usepackage{amsmath}
\usepackage{graphicx}
\usepackage{times}
\usepackage{subfigure}
\usepackage{amstext}
\usepackage{amsfonts}
\usepackage[colorlinks=true, citecolor=blue]{hyperref}
\usepackage[normalem]{ulem}

\usepackage[usenames]{color}

\providecommand{\ket}[1]{\lvert #1 \rangle}

\providecommand{\be}{\begin{equation}}
\providecommand{\ee}{\end{equation}}
\providecommand{\ba}{\begin{eqnarray}}
\providecommand{\ea}{\end{eqnarray}}

\usepackage{mathbbol}
\usepackage{mathtools}

\begin{document}
\title{Two-photon quantum Rabi model with superconducting circuits}
\author{S. Felicetti}
\address{Laboratoire Mat\' eriaux et Ph\' enom\`enes Quantiques, Sorbonne Paris Cit\' e, Universit\' e Paris Diderot, CNRS UMR 7162, 75013, Paris, France}
\author{D. Z. Rossatto}
\address{Departamento de F\'{\i}sica, Universidade Federal de S\~{a}o Carlos, 13565-905, S\~{a}o Carlos, SP, Brazil}
\author{E. Rico}
\address{Department of Physical Chemistry, University of the Basque Country UPV/EHU, Apartado 644, E-48080 Bilbao, Spain}
\address{IKERBASQUE, Basque Foundation for Science, Maria Diaz de Haro 3, 48013 Bilbao, Spain}
\author{E. Solano}
\address{Department of Physical Chemistry, University of the Basque Country UPV/EHU, Apartado 644, E-48080 Bilbao, Spain}
\address{IKERBASQUE, Basque Foundation for Science, Maria Diaz de Haro 3, 48013 Bilbao, Spain}
\address{Department of Physics, Shanghai University, 200444 Shanghai, People's Republic of China}
\author{P. Forn-D\'iaz}
\address{Barcelona Supercomputing Center (BSC), C/ Jordi Girona 29, 08034, Barcelona, Spain}

\date{\today}

\begin{abstract}
We propose a superconducting circuit to implement a two-photon quantum Rabi model in a solid-state device, where a qubit and a resonator are coupled by a two-photon interaction.
We analyze the input-output relations for this circuit in the strong coupling regime and find that fundamental quantum optical phenomena are qualitatively modified. For instance, two-photon interactions are shown to yield single- or two-photon blockade when a pumping field is either applied to the cavity mode or to the qubit, respectively.
In addition, we derive an effective Hamiltonian for perturbative ultrastrong two-photon couplings in the dispersive regime, where two-photon interactions introduce a qubit-state-dependent Kerr term. 
Finally,  we analyze the spectral collapse of the multi-qubit two-photon quantum Rabi model and find a scaling of the critical coupling with the number of qubits. Using realistic parameters with the circuit proposed, three qubits are sufficient to reach the collapse point.

\end{abstract}

\maketitle

\section{Introduction}
\label{Intro}

The realm of cavity and circuit quantum electrodynamics (QED) studies the interaction between localized  quantum-optical modes and atomic systems.
The archetypal quantum description of light-matter interaction is the Jaynes-Cummings (JC)~\cite{JaynesCummings} model, consisting of a two-level quantum emitter (qubit) interacting with a single bosonic mode. Being analytically solvable, the JC model has been ubiquitously applied to describe experiments with different quantum technologies. 
The model can be derived by first principles considering a single-electron atom in an optical cavity, applying both the dipolar and rotating-wave approximations~\cite{Milonni94}.

When the light-matter interaction strength becomes comparable with the bare frequencies of the system, the  ultrastrong-coupling (USC) regime is reached and the rotating-wave approximation (RWA) ceases to be applicable. For a qubit-cavity system in the USC regime, the JC model must be replaced with the quantum Rabi model (QRM), a non-integrable model for which an open-form analytical solution has been derived only recently~\cite{Braak2011}.
The USC regime has been experimentally achieved in different platforms, such as circuit-QED systems~\cite{Niemczyk10, FornDiaz10,Diaz2016,Pol17,Yoshihara17,Chen2017}, semiconductor quantum wells~\cite{Gunter2009, Anappara2009, Todorov09, Askenazi14, Kena-Cohen13} and terahertz metamaterials~\cite{Scalari12}. 
In particular, circuit-QED systems allow a high flexibility in the implementation of effective two-level quantum systems, namely superconducting qubits, interacting  with bosonic modes supported by microwave transmission-line and lumped-element resonators~\cite{Gu2017}. 

Together with experimental advances, the theoretical interest in the USC regime has been steadily growing, concerning fundamental properties of ultrastrong light-matter interactions~\cite{Ciuti2005,Ciuti2006,Ashhab2010, Ridolfo2012,Felicetti2014,Stassi2015,Hwang2015,Leboite2016,Rossatto2017,Garziano2015,Garziano2016,Stefano2017, Kockum2017,Stassi2017,Wang2017}, potential applications in quantum information~\cite{Rossatto2016,Nataf2011,Romero2012,Kyaw2014,Felicetti2015, Kockum17} and effective implementations~\cite{Crespi12,Ballester2012,Mezzacapo2014,Pedernales2015,Klinder2015,Puebla2016,Langford16,USCbonn,Fedortchenko17,Langford17}. Such results have prompted a number of theoretical studies on generalizations of the QRM, including multiqubit~\cite{Braak2013} and multimode~\cite{Cui2017} cases, as well as anisotropic couplings~\cite{Xie2014, Zhong2014} and two-photon interactions~\cite{Travenec2012, Albert2011,Chen2012,Duan16}. In particular,  two-photon-coupling models stand out for their highly counterintuitive spectral features. For a critical value of the coupling strength the two-photon QRM undergoes a spectral collapse~\cite{Duan16}, i.e., its discrete spectrum collapses into a continuous band. For higher values of the coupling parameter, the model becomes unbounded from below. This behavior results in a 
 nontrivial phase diagram in the many-body limit~\cite{Garbe17}.

The implementation of two-photon  couplings requires an interaction more complex than the usual dipolar case. So far, light-matter interactions beyond the dipolar approximation were for instance implemented using extremely intense optical drivings~\cite{DiPiazza2012}. 
Two-photon Lasing~\cite{Brune1987,Nikolaus1981,Gauthier1992,delValle2010,Neilinger2015} and Rabi oscillations~\cite{Bertet2002,Stufler2008,Deppe2008, Bishop2008}  have only been observed as resulting from second- or higher-order processes in resonantly-driven systems.  Quantum-simulation protocols have been proposed to effectively reproduce the two-photon QRM and observe its spectral collapse using either trapped ions~\cite{Twophot_ions,Puebla17} or cold atoms~\cite{Schneeweiss17}. However, a fundamentally nonlinear interaction is needed to observe the emergence of two-photon couplings in an undriven system.

Here, we propose a circuit-QED scheme able to implement a nondipolar ultrastrong interaction between a  flux qubit and a bosonic mode supported by a superconducting quantum interference device (SQUID). A first-principles derivation shows that exploiting the SQUID nonlinearity leads to both dipolar as well as nondipolar interaction terms, which can be selectively activated.
In particular, we focus on the case in which the dipolar interaction term is entirely suppressed, with the two-photon QRM being the main driver of the dynamics. Results of an input-output analysis show that nondipolar couplings lead to fundamentally different quantum optical properties compared to the usual dipolar case, such as the appearance of distinct selection rules and two-photon blockade as a first-order process. We also present the form of the effective Hamiltonian when both the perturbative USC and the dispersive regimes are reached. Finally, we show that the critical interaction strength to yield the spectral collapse of the of the two-photon QRM can be experimentally achieved with state-of-the-art circuit QED technology.

In Sec.~\ref{sec_scheme}, we present the circuit scheme followed by a theoretical analysis that indicates the parameter regime for which the two-photon QRM is expected to be implemented. In Sec.~\ref{sec_SC}, we focus on the strong-coupling (SC) regime of the two-photon QRM, within the validity of the RWA. We derive an input-output theory to yield the results of scattering experiments. Fundamentally different behaviors are found with respect to dipolar interactions by considering either cavity or qubit driving fields. In Sec.~\ref{sec_perturbative} we analyze the system beyond the RWA. We consider the case in which the two-photon coupling strength approaches the USC regime. We also obtain an effective Hamiltonian valid in the dispersive regime, where the frequencies of cavity and qubit are far from resonance. In Sec.~\ref{sec_collapse}, we analyze the parameter regime required to reach the spectral collapse. We show that in a feasible multiqubit configuration the collapse point corresponds to an interaction strength already achievable in state-of-the-art experiments. Furthermore, we consider higher-order correction to be added to the two-photon QRM, in order to obtain a physical model beyond the spectral collapse point. Finally, in Sec.~\ref{sec_conclusions} we summarize our results and discuss the research directions opened by the present work.

\section{Circuit scheme}
\label{sec_scheme}
Commonly in quantum optics, dipolar interactions are studied between few-level atomic-like systems coupled to other systems of bosonic nature displaying a harmonic spectrum, as for example a single mode of an electromagnetic field. The prototypical example is the JC model \cite{JaynesCummings}. The interaction operator in this case is always linear both in the atomic as well as in the harmonic system. In atomic systems, two-photon absorption can dominate over linear absorption only in presence of intense driving fields~\cite{SpectroscopyBook}. Multiphoton processes with linear interactions have been realized with the application of an external oscillating field acting at the right frequency \cite{Brune1987,Nikolaus1981,Gauthier1992,delValle2010,Neilinger2015,Bertet2002,Stufler2008,Deppe2008, Bishop2008}. 
However, an undriven system displaying a spectrum determined by intrinsic nonlinear processes requires an interaction more complex than dipolar. 
In superconducting quantum circuits, the large nonlinearity introduced by Josephson junctions naturally lead to a circuit with intrinsic nonlinear interactions. 

An interesting nonlinear circuit was already studied in the early days of quantum information with superconducting circuits when flux qubits were read out using a dc-SQUID \cite{Chiorescu2003}. Those experiments were the first to display coherent qubit-resonator oscillations by driving sideband transitions of the qubit-SQUID system \cite{Chiorescu2004}. In a separate experiment, qubit-qubit interactions were shown to be mediated by the shared SQUID detector \cite{Hime2006}. An in-depth study of the nonlinear circuit formed by qubit and SQUID using quantum network theory \cite{Burkard2004} was carried out \cite{Bertet2005b}, yielding optimal bias conditions to suppress photon-induced dephasing, which was later experimentally demonstrated \cite{Bertet2005a}. We follow here an alternative and more intuitive analysis that will aid us in understanding the physics behind the two-photon process.

\begin{figure}[]
\includegraphics[width = \columnwidth]{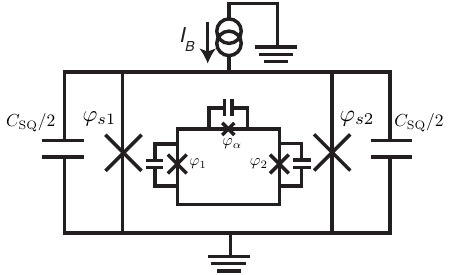}
\caption{\label{fig1} Circuit implementing the two-photon quantum Rabi model. A dc-SQUID plays the role of a nonlinear resonator inductively coupled to a flux qubit. A current bias is added to the SQUID for the sake of generality but it is not strictly necessary to implement the two-photon quantum Rabi model.}
\end{figure}

The circuit is displayed in Fig.~\ref{fig1} and it consists of a dc-SQUID with two identical junctions inductively coupled to a superconducting flux qubit. We include the possibility of current-biasing the SQUID with an external current bias line $I_B$ to keep the analysis more general. The SQUID is considered symmetric for simplicity. Any asymmetry between the SQUID junctions could always be compensated by the addition of an external optimal bias current \cite{Burkard2005}. The Hamiltonian of the SQUID circuit alone is given by 
\begin{equation}
\label{eq:H_sq}
\mathcal{\hat{H}}_{\text{SQ}}= E_C\ \hat{q}_{\text{SQ}}^2 - 2E_J \cos \left(\pi\frac{\Phi_{\text{tot}}}{\Phi_0}\right)\cos\hat{\varphi}_{\text{tot}},
\end{equation}
where $E_C=\frac{e^2}{C_{\text{SQ}}}$ is the total SQUID capacitance from Fig.~\ref{fig1}, and $E_J=I_C\Phi_0/2\pi$ is the Josephson energy of a single junction with critical current $I_C$. We defined the total SQUID capacitance $C_{\text{SQ}}$ which, as in Fig.~\ref{fig1}, may include an external shunting capacitor besides the junction self-capacitance. We denote by $\Phi_{\text{tot}}$ the total magnetic flux threading the SQUID loop. The SQUID phase $\hat{\varphi}_{\text{tot}}= \hat{\varphi}_{\text{SQ}} + \varphi_{\rm DC}$ is the sum of an externally applied constant phase $\varphi_{\rm DC}\equiv\arcsin\{ I_B/[2I_C\cos(\pi\Phi_{\rm SQ}/\Phi_0)]\}$ induced by the bias current $I_B$, plus the quantum fluctuations of the SQUID resonant mode $\hat{\varphi}_{\text{SQ}} = \hat{\varphi}_{s1}+\hat{\varphi}_{s2}$, with $\hat{\varphi}_{sk}$ being the phase operator across junction $k$, and $\hat{q}_{\text{SQ}}$ the corresponding charge operator. The phase difference $\hat{\varphi}_{s1}-\hat{\varphi}_{s2}$ is related to the external SQUID flux by fluxoid quantization, $\hat{\varphi}_{s1}-\hat{\varphi}_{s2} = 2\pi\Phi_{\text{tot}}/\Phi_0$. Here, we neglected the screening flux generated by the SQUID geometric self-inductance which is much smaller than the Josephson term in typical SQUID loop sizes of a few $\mu\rm{m}$ in length. 

Let us now consider the influence of the qubit on the SQUID. The effective magnetic dipole moment of the flux qubit generates a magnetic flux $\hat{\Phi}_{q}$ that modifies the total SQUID flux, 
$\Phi_{\text{tot}} = \hat{\Phi}_{\text{q}} + \Phi_{\text{DC}} $, where we have already considered the presence of an externally applied static flux, $\Phi_{\text{DC}}$. The qubit flux is given by $\hat{\Phi}_{\text{q}} = M\hat{I}$, with $M$ being the qubit-SQUID mutual inductance and $\hat{I}$ the current operator of the qubit. When we approximate the flux qubit as a two-level system we can simplify the current operator as $\hat{I} = I_p\hat{\sigma}_z$ \cite{Bertet2005b}, written in the basis of the persistent current states of the qubit, $\hat{\sigma}_z = |L\rangle\langle L| - |R\rangle\langle R|$, with $|L\rangle$ and $|R\rangle$ being the left- and right-circulating persistent current states, respectively. We will consider that the flux generated by the qubit is small $\langle\hat{\Phi}_{\text{q}}\rangle/\Phi_0\ll1$. If we further assume the SQUID to be in the phase regime $E_J/E_C\gg1$, the values of the SQUID phase operator near its ground state are small $|\hat{\varphi}_{\text{SQ}}|\ll\pi$ \cite{Koch2007}. This approximation is justified when we consider realistic SQUID resonance frequencies in the range of a few GHz, in proximity to the qubit resonance. Under these conditions, the SQUID Josephson potential can be doubly expanded leading to a collection of terms that contain all physics of the qubit-SQUID interaction to all orders. Up to second order, we have
\begin{multline}\label{eq:Usqapp}
U_{\text{SQ}}\simeq-2E_J\left[\cos\left(\pi\frac{\Phi_{\text{DC}}}{\Phi_0}\right) - \frac{\pi}{\Phi_0}\sin\left(\pi\frac{\Phi_{\text{DC}}}{\Phi_0}\right)\hat{\Phi}_{\text{q}} \right]\times\\ \left[\cos(\varphi_{\text{DC}}) -\sin(\varphi_{\text{DC}})\hat{\varphi}_{\text{SQ}} - (1/2)\cos(\varphi_{\text{DC}})\hat{\varphi}_{\text{SQ}}^2))\right].
\end{multline}
Here, $\Phi_{\text{DC}}$ and $\varphi_{\text{DC}}$ correspond to the constant flux and current bias, respectively, and they can be independently tuned. Using the harmonic oscillator basis for the SQUID mode, we can now express the SQUID phase operator with annihilation (creation) operator $\hat{a}$ $(\hat{a}^{\dag})$
\begin{equation}
\hat{\varphi}_{\text{SQ}} = \frac{2\pi}{\Phi_0}\sqrt{\frac{\hbar\omega_{\text{SQ}}L_J}{2}}(\hat{a}+\hat{a}^{\dag}),
\end{equation}
where we defined the SQUID resonance frequency $\omega_{\text{SQ}}\equiv(L_JC_{\text{SQ}})^{1/2}$. The inductance $L_J$ appearing here is the Josephson inductance of the SQUID. Its explicit form can be found by considering the term proportional to $\hat{\varphi}^2_{\text{SQ}}$ from Eq.~(\ref{eq:Usqapp}), $E_J\cos(\pi\Phi_{\text{DC}}/\Phi_0)\cos(\varphi_{\text{DC}})\hat{\varphi}^2_{\text{SQ}}$. Converting the phase operator to flux operator with $\hat{\varphi}_{\text{SQ}} = (2\pi/\Phi_0)\hat{\Phi}$, we obtain an inductive-like potential energy
\begin{equation}
U_J = \frac{2\pi I_C}{\Phi_0}\cos(\pi\Phi_{\text{DC}}/\Phi_0)\cos(\varphi_{\text{DC}})\hat{\Phi}^2\equiv\frac{\hat{\Phi}^2}{2L_J(\Phi_{\text{DC}},\varphi_{\text{DC}})},
\end{equation}
where the Josephson inductance of the SQUID has been defined 
\begin{equation}\label{eq:sq_lj}
L_J(\Phi_{\text{DC}},\varphi_{\text{DC}})\equiv\frac{\Phi_0}{2\pi (2I_C)\cos(\pi\Phi_{\text{DC}}/\Phi_0)\cos(\varphi_{\text{DC}})}.
\end{equation}

The first-order interaction term in Eq.~(\ref{eq:Usqapp}) leads to the usual Jaynes-Cummings model,
\begin{multline}
U_{\text{JC}} = -4E_J\left(\frac{\pi}{\Phi_0}\right)^2\sin\left(\pi\frac{\Phi_{\text{DC}}}{\Phi_0}\right)\sin(\varphi_{\text{DC}})\times\\MI_p\sqrt{\frac{\hbar\omega_{\text{SQ}}L}{2}}(\hat{a}+\hat{a}^{\dag})\hat{\sigma}_z.
\end{multline}
Notice that this term will vanish when no current biases the SQUID, $\varphi_{\text{DC}} = 0$. This cancellation takes place since the SQUID was assumed symmetric and therefore to first order the external current generates no net flux in the qubit loop. The difference with the usual linear coupling to a resonator is due to the interaction not being dipolar but mediated by the $\cos(\pi\Phi_{\text{tot}}/\Phi_0)$ factor in Eq.~(\ref{eq:H_sq}). The next order term in the expansion of Eq.~(\ref{eq:Usqapp}) is the one we are really interested in as it leads to the two-photon Rabi (TPR) physics:
\begin{multline}
U_{\text{TPR}} = -4E_J\left(\frac{\pi}{\Phi_0}\right)^3\sin\left(\pi\frac{\Phi_{\text{DC}}}{\Phi_0}\right)\cos(\varphi_{\text{DC}})\times\\MI_p\frac{\hbar\omega_{\text{SQ}}L}{2}(\hat{a} + \hat{a}^{\dag})^2\hat{\sigma}_z.
\end{multline}
Using no bias current $\varphi_{\text{DC}} = 0$
\begin{equation}\label{eq:utpr}
U_{\text{TPR}} = -\frac{\pi}{4}\tan\left(\pi\frac{\Phi_{\text{DC}}}{\Phi_0}\right)\frac{MI_p}{\Phi_0}\hbar\omega_{\text{SQ}}(\hat{a} + \hat{a}^{\dag})^2\hat{\sigma}_z.
\end{equation}
This interaction term was also derived using other methods by Bertet \emph{et al.}~\cite{Bertet2005b}, in order to analyze the dephasing of flux qubits due to thermal fluctuations. Notice that, as already mentioned, the qubit operator is written in the flux basis. When transforming to the energy basis and biasing the qubit at the symmetry point, the coupling operator $\hat{\sigma}_z$ transforms into $\hat{\sigma}_x$. It is clear from Eq.~(\ref{eq:utpr}) that the DC-flux in the SQUID ($\Phi_{\text{DC}}$) needs to be different from 0 to switch ON this interaction. In other words, it is tunable. Note that the ``spurious" terms proportional to $\hat{a}^{\dag}\hat{a}+\hat{a}\hat{a}^{\dag}$ can be re-arranged as $\sim\hbar\omega_{\rm{QS}}(2\hat{a}^{\dag}\hat{a}+1)\hat{\sigma}_z$, which directly add to the frequency of the SQUID resonance mode and depend on the qubit state. Therefore (if we neglect the pure 2-photon terms $\sim \hat{a}^2+\text{H.c.})$ we have the Hamiltonian equivalent to the usual AC-Stark shift used in conventional circuit-QED settings but here we do not require the qubit and SQUID resonances to be detuned. This process can be regarded as an additional inductance coming from the qubit which is quantized and can change sign. Therefore, the qubit state can be directly read out by measuring the resonance frequency of the SQUID. Such a technique was already implemented in the bifurcation readout method \cite{Lupascu2007, deGroot2010}.

We can now define a two-photon coupling strength as $\hbar g_2 \equiv -(\pi/4)\tan\left(\pi\Phi_{\text{DC}}/\Phi_0\right)(MI_p/\Phi_0)\hbar\omega_{\text{SQ}}$. The entire circuit Hamiltonian using the two-level approximation for the flux qubit reads
\begin{equation}
\mathcal{\hat{H}} = \hbar\omega_{\text{SQ}}\left(\hat{a}^{\dag}\hat{a}+\frac{1}{2}\right) + \frac{\hbar\omega_{\text{q}}}{2}\hat{\sigma}_z + \hbar g_2(\hat{a}+\hat{a}^{\dag})^2\hat{\sigma}_x,
\label{original2phHam}
\end{equation}
which is the canonical two-photon quantum Rabi model. 

Using the expansion of the SQUID potential, we can calculate the next-order interaction term. This will be important to study the response of the system in the regime where the potential is unbounded from below (see Sec.~{\ref{sec_collapse}}). The next term in the expansion is fourth order in the SQUID photon operator
\begin{multline}
U_{\text{4P}} = 2E_J\frac{\pi}{\Phi_0}\sin\left(\pi\frac{\Phi_{\text{DC}}}{\Phi_0}\right)\hat{\Phi}_{\text{q}}\frac{1}{4!}\cos(\varphi_{\text{DC}})\hat{\varphi}^4 \\=\frac{E_J}{12}\frac{\pi}{\Phi_0}\sin\left(\pi\frac{\Phi_{\text{DC}}}{\Phi_0}\right)\cos(\varphi_{\text{DC}})\left(\frac{2\pi}{\Phi_0}\right)^4\times\\\left(\frac{\hbar\omega_{\text{SQ}}L_J}{2}\right)^2MI_p\hat{\sigma}_z(\hat{a} + \hat{a}^{\dag})^4.
\end{multline}
Notice that this term has the opposite sign compared to the second-order one. We can rewrite the fourth-order term using the explicit form of the SQUID inductance in Eq.~(\ref{eq:sq_lj}),
\begin{multline}
U_{\text{4P}} = \frac{1}{96}\left(\frac{\pi}{\Phi_0}\right)^2\frac{\tan(\pi\Phi_{\text{DC}}/\Phi_0)}{\cos(\pi\Phi_{\text{DC}}/\Phi_0)}\frac{(\hbar\omega_{\text{SQ}})^2}{I_C\cos\varphi_{\text{DC}}}\times \\MI_p\hat{\sigma}_z(\hat{a} + \hat{a}^{\dag})^4.
\end{multline}
Relative to the second-order term,
\begin{equation}
\frac{|U_{\text{4P}}|}{|U_{\text{TRP}}|} = \frac{1}{24}\frac{\pi}{\Phi_0}\frac{\hbar\omega_{\text{SQ}}}{I_C\cos(\pi\Phi_{\text{DC}}/\Phi_0)}.
\label{quarticTerm}
\end{equation}
Assuming a typical order of magnitude of the SQUID current of $I_C\sim1\mu$A, $\Phi_0I_C\sim3$~THz and a SQUID resonance of 5~GHz, then $|U_{\text{4P}}|/|U_{\text{TPR}}|\sim10^{-3}$. Therefore, the circuit in Fig.~\ref{fig1} when no current is applied to the SQUID, $I_B=0$, gives us a very good approximation of a genuine two-photon QRM, under the approximations assumed throughout this section.

\section{Strong-coupling regime}
\label{sec_SC}
\begin{figure}[]
\centering
\includegraphics[angle=0, width=0.5\textwidth]{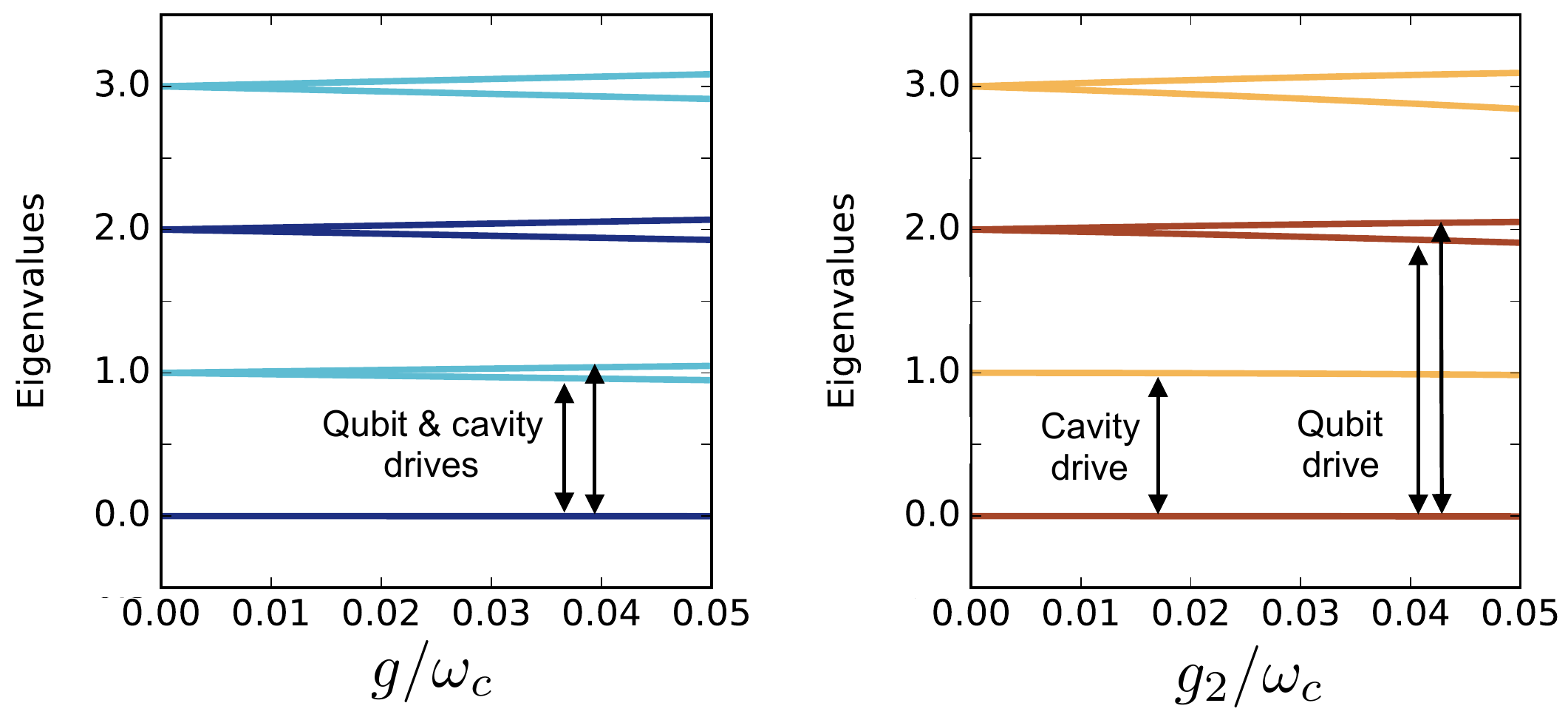}
\caption{\label{SCspectrum}  
 Comparison of the eigenenergies of the standard one-photon and the two-photon QRMs in the strong-coupling regime, as a function of the coupling strength $g$. In both cases, we consider resonant interactions, so that $\omega_q = \omega_c$ for dipolar coupling (left), while $\omega_q =2 \omega_c$  in the two-photon case (right). Lighter and darker colors in each plot identify parity eigenstates, corresponding to the operators $\Pi_{\text{JC}}=\sigma_z e^{i\pi a^\dagger a}$ and $\Pi_{\text{2ph}}=e^{i\pi a^\dagger a}$ for the one- and two-photon QRM, respectively. Vertical arrows show selection rules for cavity and qubit drivings (see Sec.~\ref{subsec:inout}).}
\end{figure}

Having obtained a circuit scheme that implements a cavity-QED system with an intrinsic nonlinear coupling, we  characterize in this section the quantum-optical properties of such a model by analyzing the spectral features of the system. We also derive an input-output theory to understand the response the system would have in actual scattering experiments. A comparison with the standard one-photon QRM shows that replacing dipolar couplings with two-photon interactions strongly modifies fundamental features of qubit-cavity systems, leading to nontrivial selection rules and multiphoton blockade.  
This section focuses on the strong-coupling (SC) regime, in which the coupling strength is small compared to both the qubit and cavity bare frequencies, but it is still larger than all dissipation rates. Notice that the coupling strength in the proposed circuit implementation is tunable [Eq.~(\ref{eq:utpr})], therefore a wide region of parameters could be exploited with a single device.
Hence, the SC represents the most natural regime to begin with in a practical implementation.

Let us briefly discuss the Hamiltonian spectrum, which will be needed to understand the dynamical response of the system to external drivings. In the SC regime, the two-photon QRM Hamiltonian can be simplified by RWA to the  two-photon JC Hamiltonian (we omit the hat symbol over operators in the rest of the article),
\begin{equation}
 \mathcal{H}_{\text{RWA}} = \omega_c  a^\dagger  a + \frac{\omega_q}{2} \sigma_z + g_{2} \! \left( \sigma_+  a^2 +  \sigma_- {a^\dagger}^2 \right),
\label{2phSC}
\end{equation}
where we neglected the counter-rotating terms $\sigma_- a^2 + \text{H.c.}$ and $ \sigma_x a^\dagger a$, rotating at frequencies $2\omega_q+\omega_c$ and $\omega_q$, respectively. Here $\sigma_\pm = \frac{1}{2} (\sigma_x \pm i\sigma_y)$. Fig.~\ref{SCspectrum} shows a comparison between the spectrum of the two-photon QRM and the standard single-photon QRM, which reduces to the JC model in the SC regime
\begin{equation}
 \mathcal{H}_{\text{JC}} = \omega_c  a^\dagger  a + \frac{\omega_q}{2} \sigma_z + g\left( \sigma_+  a +  \sigma_- {a^\dagger} \right).
\label{JC}
\end{equation}
We consider both models for resonant interactions, so that in the two-photon case we set $\omega_q = 2\omega_c$, while for the one-photon coupling the qubit energy spacing is equal to the cavity frequency. These are the most natural frequency scales to compare the two models, as in both cases the so-called vacuum Rabi splitting is maximal.

In the SC regime, both models respect a continuous symmetry given by the total excitation number for dipolar interactions $C_{\text{JC}} = a^\dagger a + \sigma_z$, and a weighted excitation number for two-photon couplings $C_{\text{2ph}} =2 a^\dagger a + \sigma_z$. This symmetry leads to an exact solution of the JC model \cite{Klimov09} and its two-photon generalization \cite{Sukumar1981}.
For resonant two-photon coupling, the ground and first-excited states are given by $\ket{\text{g},0}$ and $\ket{\text{g},1}$, which are separated by $\omega_{c}$. Here, $\ket{\text{g}}$ ($\ket{\text{e}}$) denotes the qubit ground (excited) state while the field state is represented in the basis of Fock states. Higher excited states are given by the doublets $\ket{\psi_n^\pm}=\left( \ket{\text{g}, n+2} \pm \ket{\text{e},n} \right)/\sqrt{2}$. Setting the ground-state energy to zero, the energy eigenvalues corresponding to the doublets are 
$E_n^\pm= \omega_{c} (n+2) \pm g_{2}\sqrt{(n+1)(n+2)}$. 

When the coupling strength is too large to allow the RWA, the continuous symmetries $C_{\text{JC}}$ and $C_{\text{2ph}}$ break down into discrete ones, which we denote by $\Pi_{\text{JC}}$ and $\Pi_{\text{2ph}}$
for dipolar and two-photon couplings, respectively.  The standard QRM preserves the  parity of the total number of excitations $\Pi_{\text{JC}}=\sigma_z e^{i\pi a^\dagger a}$. On the other hand, the two-photon coupling Hamiltonian of Eq.~\eqref{original2phHam} commutes with the parity of the photon number $\Pi_{\text{2ph}}=e^{i\pi a^\dagger a}$. The latter symmetry is unaffected by spin-flips, hence it is conserved and it introduces selection rules in the case of qubit driving, as detailed in the following. Notice that the symmetry is different when considering a pure two-photon coupling $a^2 + (a^{\dag})^2$ rather than a full quadratic coupling $(a+a^{\dag})^2$ (see Sec.~\ref{sec_collapse}).

\subsection{Input-output theory}
\label{subsec:inout}
In this section we analyze the photon statistics of the cavity output field for the driven two-photon QRM, and compare it to the results of the one-photon QRM. We assume that the system is coupled to two waveguides that support a continuum of modes. We consider the cases in which either the cavity or the qubit is coupled to an input waveguide. In both configurations we observe the state emitted by the cavity into an output waveguide. We will focus on the stationary output field $a_{\text{out}}= \lim_{t\rightarrow \infty} \displaystyle a_{\text{out}}(t)$, obtained after an evolution time that is long compared to the relaxation time of the system.

In the following, we present results on the system output obtained by numerically solving the master equation (see Appendix~\ref{inout}).
Input-output relations~\cite{Walls94} are obtained under the RWA on the system-bath coupling, assuming that the interaction strength with the extracavity mode is weak and constant over the relevant frequency range. Notice that these approximations are well justified only when $g,g_{2}\ll \omega_c, \omega_q$. When the USC regime is reached, the input-output relations must be modified by expressing the electric-field operator in the cavity-qubit dressed basis~\cite{Ridolfo2012,Leboite2016}, or by taking into account the colored nature of the dissipation bath~\cite{Ciuti2006,Liberato2009}.

In particular, we consider the normalized transmitted intensity $T= n_{\text{out}} /n_{\text{in}} $, and the two-photon correlation function  
$g^{(2)}(0) =   \langle (a_{\text{out}}^\dagger)^{2} (a_{\text{out}})^{2} \rangle/n_{\text{out}}^2$, with $n_{\text{out}} = \langle a_{\text{out}}^\dagger a_{\text{out}} \rangle$. The transmitted state is analyzed as a function of the frequency $\omega_d$ and intensity $D$ of the coherent driving field (see Appendix~\ref{inout}).

\subsubsection{Cavity driving}
\begin{figure}[]
\centering
\includegraphics[angle=0, width=0.5\textwidth]{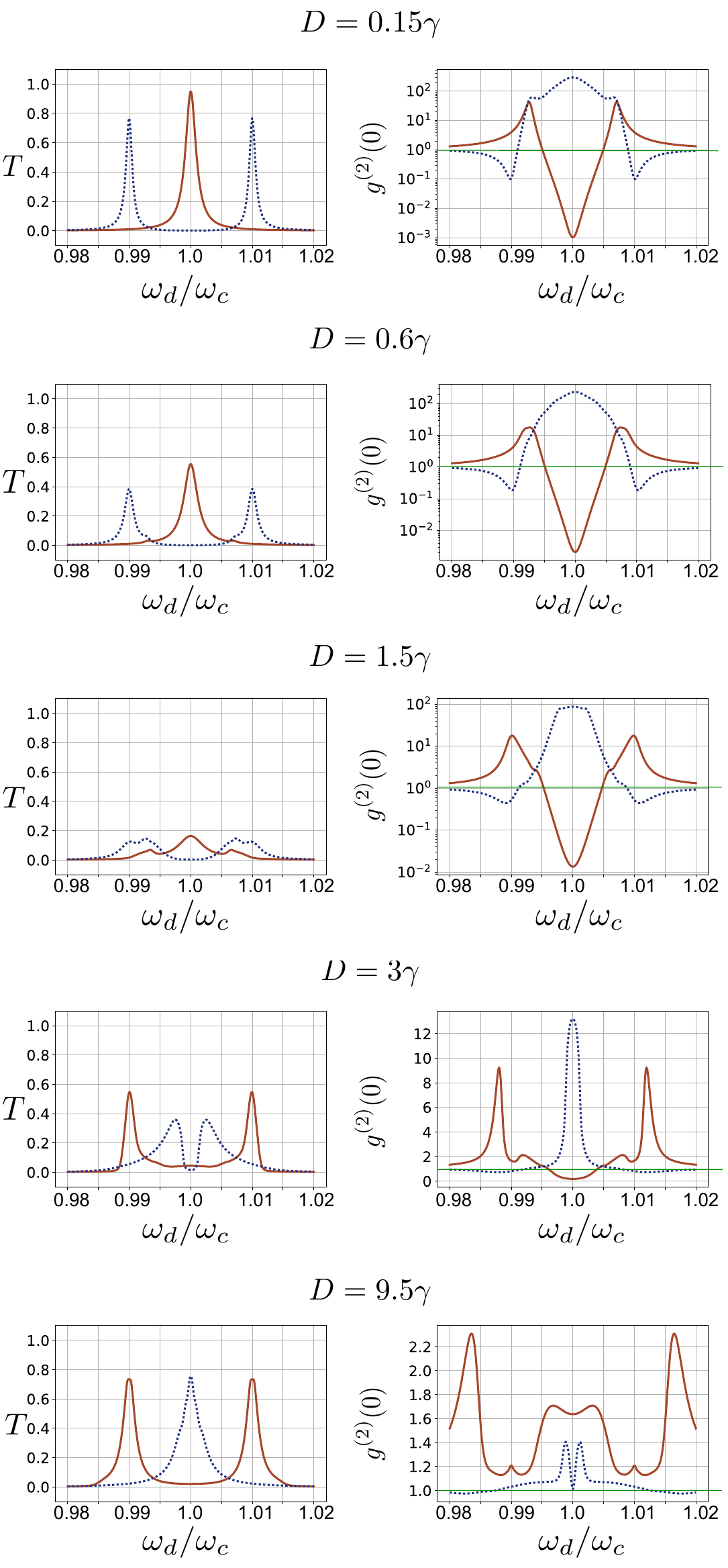}
\caption{\label{cavityDrive} System output spectrum for coherent cavity driving. One-photon (blue dashed line) and two-photon (red solid line) interactions are compared. The plots show normalized transmitted intensity (left column) and two-photon correlation function (right column) as a function of the driving frequency $\omega_d$. From the top to the bottom, the intensity $D$ of the coherent driving is increased (notice that $(D/\gamma)^2$ corresponds to the intracavity photons in an empty resonator driven by the same field). The horizontal green line in the plots of the right column marks the transition from super-poissonian $ g^{(2)}(0)>1$ to sub-poissonian $g^{(2)}(0)<1$ photon statistics.  The qubit-cavity coupling is $g=g_{2}=0.01\omega_c$, the cavity decay rate is
$\gamma=\omega_c\times10^{-3}$, while the qubit decay and pure dephasing rates are given by $\gamma_q=\omega_c\times10^{-4}$ and
$\gamma_\phi=5\omega_c\times10^{-5}$, respectively.}
\end{figure}

Let us first consider the case in which the cavity mode is continuously driven by an input coherent state.
In Fig.~\ref{cavityDrive}, we compare the results obtained from the two-photon quantum Rabi model (red solid line) and from the standard one-photon interaction (blue dashed line). The introduction of two-photon interactions in a qubit-cavity system immediately results in fundamental modifications of the cavity transmission properties.
For a weak input field the one-photon model is characterized by the vacuum Rabi splitting, visible in the symmetric transmission peaks at frequencies given by $\omega_c \pm g$. On the other hand, the transmitted intensity of the two-photon QRM presents a single peak at the cavity frequency.  

The differences in the response of the two models can be understood by comparing their energy spectra (see Fig.~\ref{SCspectrum}). For dipolar interaction, the first allowed transitions lead to the excitation of the lowest JC doublets $\ket{\text{g},0}\pm \ket{\text{e},1}$. The first excited state of the two-photon QRM in the SC regime is instead given by $\ket{\text{g},1}$, which contains no atomic excitations. As a result, the weakly-driven transmitted intensity is the same than for an empty cavity. However, the presence of the nonlinearity in the system is revealed by the dip of the two-photon correlation function at resonance, as the two- and higher-order photon components in the transmitted state are strongly damped (sub-Poissonian statistics). Hence, although the transmission profile is similar to the empty-cavity case, the transmitted photons are strongly antibunched, such that the system exhibits single-photon blockade~\cite{Imamoglu97}. This blockade means that a single-photon absorption event prevents the system from absorbing further photons, resulting in an emission of photons one by one. Therefore, in this regime the two-photon QRM represents a steady-state source of single-photon states.

When the amplitude of the input field increases, the ratio of transmitted field in the two-photon QRM substantially decreases. High transmission is restored in the limit when the amplitude of the driving field is large compared with the dissipation rates. In that case, qubit saturates in both models and the system asymptotically reaches a harmonic behavior. The one-photon coupling model has an almost linear response, i.e., it shows a single high-transmission peak at the cavity frequency, with  $g^{(2)}(0)\sim 1$.  On the other hand, in the two-photon QRM the presence of the qubit strongly affects the transmission spectrum in the strong-driving limit. In this case, the output spectrum is characterized by two transmission peaks, detuned from the cavity frequency by $\pm g_{2}$, a frequency gap not corresponding to the two-photon Rabi splitting $\sqrt{2}g_{2}$. Also, for these peaks the two-photon correlation function is almost unitary. In Appendix~\ref{StongDrive} we provide an analytical explanation for this behavior, assuming that for a strong driving the steady-state of the system is characterized by highly-excited eigenstates.

\subsubsection{Qubit driving}
\begin{figure}[]
\centering
\includegraphics[angle=0, width=0.45\textwidth]{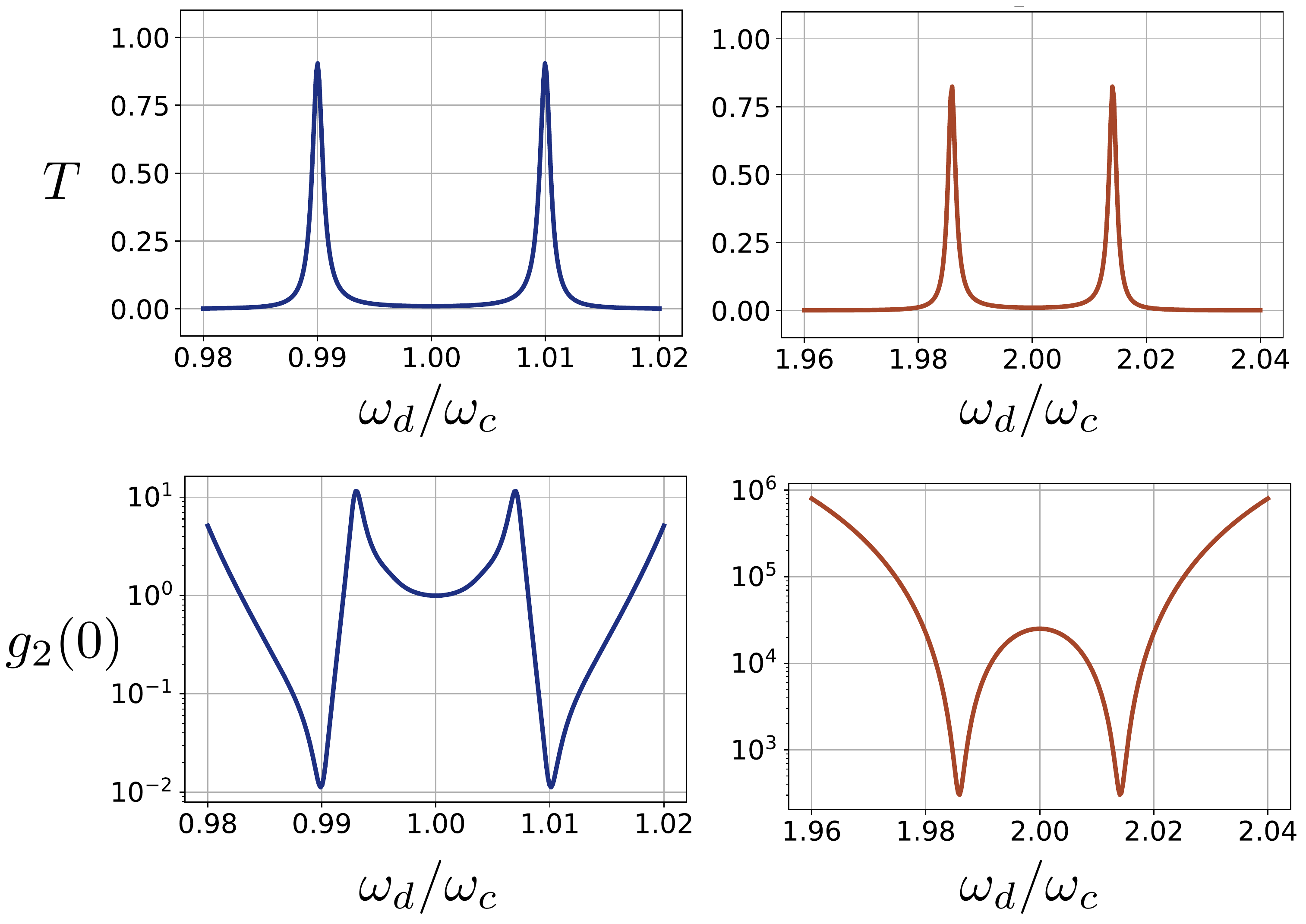}
\caption{\label{qubitDrive} System transmission spectrum for coherent qubit driving. The plots show the normalized transmitted intensity $T$ and the two-photon correlation function $g^{(2)}(0)$, for the standard one-photon QRM (left column) and the two-photon QRM (right column). Notice that for two-photon couplings the resonances are centered around $2\omega_c$ and the photon statistics is strongly super-Poissonian. The following physical parameters have been used: driving intensity $D=0.03 \gamma$, coupling strength $g,g_{2}=0.01\omega_c$, cavity and qubit decay rates $\gamma=\gamma_q=\omega_c\times 10^{-3} $, and qubit pure dephasing rate $\gamma_\phi = 5\omega_c\times 10^{-5}$.}
\end{figure}

The transmission spectrum when the qubit is driven by a continuous coherent field is shown in Fig.~\ref{qubitDrive}. The spectrum is characterized by a double-peak structure which corresponds to the vacuum Rabi splitting, respectively equal to $2g$ and to $2\sqrt{2} g_{2}$ for the one- and two-photon QRM. Notice the driving frequency of the two-photon QRM spectrum being twice the frequency of the cavity. Indeed, transitions between the ground and the  first-excited state are forbidden by selection rules (see Fig.~\ref{SCspectrum}). The two-photon QRM Hamiltonian \textendash in all regimes of interaction \textendash commutes with the symmetry operator $\Pi_{\text{2ph}} = \exp\{ i\pi a^\dagger a \}$, which corresponds to the parity of the photon number. Unlike the case of the driven cavity, the qubit driving does not break this symmetry, so that transitions between states with opposite parity are strongly damped.

A marked difference between one- and two-photon QRMs is also found for the photon statistics of the cavity transmitted field. In the one-photon case we observe sub-Poissonian statistics ($g^{(2)}(0)<1$) at each  resonance of the first JC doublet, indicating a suppression of multiple excitations which is indeed due to a single-photon blockade~\cite{Hamsen2017}. On the other hand, the $g^{(2)}(0)$ function of the two-photon QRM is always larger than 1 (super-Poissonian statistics), which is an indication of a higher-photon number in the cavity output field. Therefore, the two-photon QRM does not exhibit single-photon blockade when a driving field is applied to the qubit. The next step is to analyze if two-photon blockade exists.

\begin{figure}[]
\centering
\includegraphics[angle=0, width=0.5\textwidth]{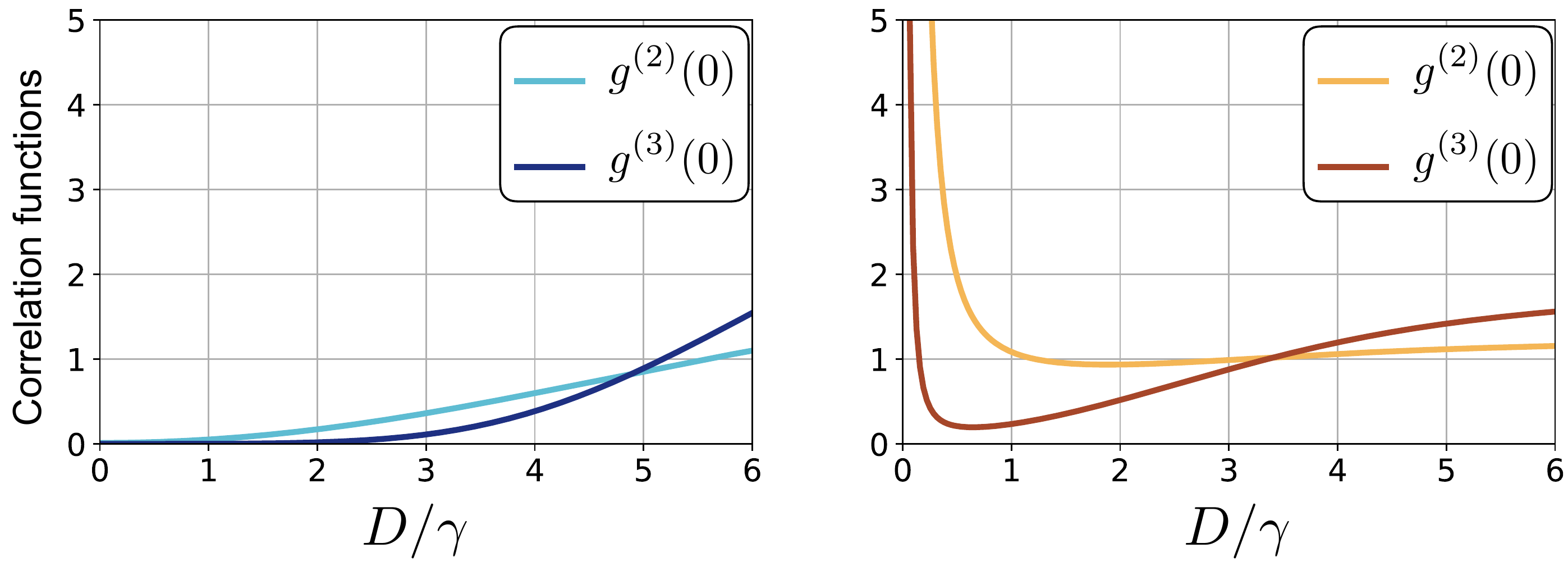}
\caption{\label{g2g3}  Two- and three-photon correlation function as a function of the driving intensity $D$ in the case of qubit driving. For the dipolar QRM (left), we observe photon antibunching for low driving intensity, due to the spectrum nonlinearity. For the two-photon QRM (right) there is an easily-accessible region of parameters in which the two-photon correlation function is larger than or close to 1, while three-photon contributions are strongly dumped. This quantum nonlinear effect is known as two-photon blockade. The frequency of the driving frequency is taken to match the high-transmission peaks (see Fig.~\ref{qubitDrive}). Accordingly, $\omega_d = \omega_c + g$ in the left panel, while $\omega_d = 2\omega_c + \sqrt{2}g_2$ in the right panel. The following physical parameters have been considered: coupling strength $g,g_{2}=0.01\omega_c$, cavity and qubit decay rates $\gamma=\gamma_q=\omega_c\times 10^{-3} $, and qubit pure dephasing rate $\gamma_\phi = 5\omega_c\times 10^{-5}$.}
\end{figure}

In order to provide evidence for two-photon blockade when $n_{\text{out}}/\gamma \ll 1$ (very small average number of photons in the intracavity field), it is sufficient to simultaneously fulfill $g^{(2)}(0) \ge 1$ and $g^{(3)}(0)= \langle (a_{\text{out}}^\dagger)^3 (a_{\text{out}})^3 \rangle/n_{\text{out}}^3<1$ (three-photon correlation function) \cite{Hamsen2017}. These conditions indicate two-photon bunching and three-photon antibunching, respectively. In Fig.~\ref{g2g3} we show the two- and three-photon correlation functions as the driving intensity $D$ is increased, fixing the driving frequency $\omega_d$ to the values that maximize transmission for one- and two-photon interactions [see Fig.~\ref{qubitDrive}]. Notice that the two-photon QRM exhibits a wide region of values of $D$ for which simultaneously $g^{(2)}(0)\ge1$ and $g^{(3)}(0) <1$, signaling the presence of two-photon blockade, i.e., the absorption of two photons by the cavity (via the absorption of a single photon by the qubit) prevents the system from absorbing further photons. 
Such a dynamical quantum-nonlinear effect has been observed in an atom-cavity system (JC system) driven close to a two-photon resonance~\cite{Hamsen2017}. The circuit scheme here proposed allows the implementation of this phenomenon as a first-order process, on a platform capable of reaching the USC regime (see Sec.~\ref{sec_perturbative}). 

\section{Perturbative Ultrastrong Coupling}
\label{sec_perturbative}

In this section, we consider the first correction to the RWA Hamiltonian, Eq.~(\ref{2phSC}), when the coupling strength approaches the ultrastrong coupling (USC) regime. 
\subsection{Two-photon Bloch-Siegert Hamiltonian}
In Sec.~\ref{sec_SC}, the counter-rotating terms ($\sigma_- a^2 +$ H.c. and $\sigma_{x}a^\dagger a$) were neglected to obtain the two-photon QRM in the strong coupling regime [Eq.~(\ref{original2phHam})]. Instead, these terms can be treated in a perturbative fashion 
in the perturbative USC regime \cite{Rossatto2017}, i.e. when $g_{2}(\bar{n}+1) \ll \omega_{q} , 2\omega_{c}+\omega_{q}$, with $\bar{n}$ being the average photon number in the cavity mode. By applying a perturbation theory on the counter-rotating terms up to second order in $g_{2}/( 2\omega_{c}+\omega_{q})$ and $g_{2}/\omega_{q}$, we obtain the two-photon Bloch-Siegert (2BS) Hamiltonian (neglecting constant terms),
\begin{align}
\mathcal{H}_{\text{2BS}} &= \mathcal{H}_{\text{RWA}} - \omega_{\text{2BS}}a^{\dagger}a + (\omega_{\text{2BS}} + \Omega_{q})\frac{\sigma_{z}}{2} \nonumber \\
&+ \left( \frac{\omega_{\text{2BS}}}{2} + 2\Omega_{q} \right)\! \sigma_{z} \! \left[a^{\dagger}a + (a^{\dagger}a)^{2} \right], \label{2phBS}
\end{align}
where $\omega_{\text{2BS}} = 2g_{2}^{2}/(2\omega_{c}+\omega_{q})$ and $\Omega_{q} = 2g_{2}^{2}/\omega_{q}$ are the two-photon Bloch-Siegert (BS) shifts due to the counter-rotating terms.

It is interesting to compare $\mathcal{H}_{\text{2BS}}$ with the BS Hamiltonian derived for the standard one-photon QRM \cite{Klimov09,Rossatto2017}
\begin{equation}
\mathcal{H}_{\text{BS}} = \mathcal{H}_{\text{JC}} + \omega_{\text{BS}} \frac{\sigma_{z}}{2} + \omega_{\text{BS}}\sigma_{z}a^{\dagger}a, \label{1phBS}
\end{equation}
with $\omega_{\text{BS}} = g^{2}/(\omega_{c}+\omega_{q})$ the Bloch-Siegert shift \cite{FornDiaz10}. First, we observe a natural and expected difference between $\omega_{\text{2BS}}$ and $\omega_{\text{BS}}$, in the prefactor $2$ of the $\omega_c$ contribution to $\omega_{\text{2BS}}$. Then, aside from the BS shift proportional to $\Omega_{q}$ due to the cross term $\sigma_{x}a^{\dagger}a$, which is exclusive to the two-photon QRM, we note two additional features in the two-photon BS Hamiltonian that are not present in the one-photon case Eq.~(\ref{1phBS}), up to the second order in $g$.  Namely, a negative shift in the cavity frequency ($ - \omega_{\text{2BS}}a^{\dagger}a$) and the presence of a nonlinearity in the cavity mode conditioned on the qubit state [$\propto \sigma_{z}(a^{\dagger}a)^{2}$].

With $\mathcal{H}_{\text{2BS}}$, which is diagonalizable, one could extend the analysis of the dissipative dynamics performed in Sec.~\ref{sec_SC} into the pertubative USC regime. In that case, it is necessary to replace the phenomenological optical master equation (Appendix \ref{inout}) by the Bloch-Redfield (or dressed-state) master equation \cite{beaudoin2011}. Moreover, one has to consider the input-output relations that express the electric-field operator in the cavity-qubit dressed basis~\cite{Ridolfo2012,Leboite2016}, and take into account the colored nature of the dissipation bath~\cite{Ciuti2006,Liberato2009}.

\subsection{Two-photon dispersive regime}
In this section, we consider the case in which the frequencies of cavity and qubit are far from resonance (known as the dispersive regime). This configuration is widely used for qubit readout in circuit QED \cite{Schuster2005}, where the qubit-cavity interaction is given by the standard one-photon QRM.

In this regime, achieved when $|\Delta_{\text{2ph}}|=|2\omega_{c} - \omega_{q}| \gg g_{2}(\bar{n}+1)$, both rotating and counter-rotating terms perturbatively contribute to the dynamics. In this way, by applying a perturbation theory on the entire interaction Hamiltonian of Eq.~\eqref{original2phHam} up to second order in $g_{2}/( 2\omega_{c}+\omega_{q})$, $g_{2}/\omega_{q}$ and $g_{2}/|\Delta_{\text{2ph}}|$, we obtain the two-photon QRM Hamiltonian in the dispersive regime (neglecting constant terms)
\begin{align}
\mathcal{H}_{\text{dis}} &=  (\omega_{c} - \omega_{\text{2BS}}+\chi)a^{\dagger}a \nonumber \\
&+ (\omega_{q}+ \omega_{\text{2BS}} + \Omega_{q} + \chi)\frac{\sigma_{z}}{2} \nonumber \\
&+ \left( \frac{\omega_{\text{2BS}}}{2} + 2\Omega_{q} + \frac{\chi}{2}\right)\! \sigma_{z} \! \left[a^{\dagger}a + (a^{\dagger}a)^{2} \right], \label{2disp}
\end{align}
with $\chi = 2g_{2}^{2}/\Delta_{\text{2ph}}$. When the counter-rotating terms are negligible ($\omega_{\text{2BS}}, \Omega_{q} \to 0$), we end up with the two-photon JC model in the dispersive regime
\begin{align}
\mathcal{H}^{\text{RWA}}_{\text{dis}} &=  (\omega_{c}+\chi)a^{\dagger}a + (\omega_{q} + \chi)\frac{\sigma_{z}}{2} \nonumber \\
&+ \frac{\chi}{2} \sigma_{z} \! \left[a^{\dagger}a + (a^{\dagger}a)^{2} \right]. \label{2dispJC}
\end{align}

It is interesting to compare $\mathcal{H}^{\text{RWA}}_{\text{dis}}$ with the dispersive regime of the standard one-photon Jaynes-Cummings model. Up to fourth order in $g/|\omega_c - \omega_q|$, the dispersive JC model is given by
\begin{align}
\mathcal{H}^{\text{JC}}_{\text{dis}} &=  (\omega_{c}+\zeta)a^{\dagger}a + (\omega_{q} + \chi^{(1)})\frac{\sigma_{z}}{2} \nonumber \\
&+ \chi^{(1)} \sigma_{z} a^{\dagger}a + \zeta \sigma_{z}(a^{\dagger}a)^{2}, \label{dispJC}
\end{align}
in which $\chi^{(1)} = g^{2}/\Delta$ is the qubit-cavity dispersive coupling, the AC-Stark shift, with $\Delta = \omega_{c} - \omega_{q}$ ($|\Delta| \gg g)$, and $\zeta = g^{4}/\Delta^{3}$ is a small nonlinearity that is usually neglected \cite{beaudoin2011}.

Let us briefly comment on the four terms introduced by the two-photon dispersive interaction. In the first line of Eq.~\eqref{2dispJC} we have an identical constant frequency shift $\chi$ for both the qubit and cavity mode. This shift is the equivalent of the Lamb shift in the dispersive JC model [first line of Eq.~\eqref{dispJC}]. In the JC model, the shift in the qubit frequency is due to a perturbation of second order in $g$, while the shift in the cavity frequency $\zeta$ comes from a perturbation of fourth order in $g$. The first term in the second line of  Eq.~\eqref{2dispJC} is the analogous of the AC-Stark effect, i.e., a shift $\chi$ of the cavity mode frequency conditioned on the qubit state. The last term in Eq.~(\ref{2dispJC}) is a more interesting one since it consists in a nonlinear Kerr effect that depends on the qubit state. However, in the two-photon QRM this nonlinearity is not due to a small higher-order correction as occurs in the one-photon QRM [last term in Eq.~(\ref{dispJC})], but it is of the same order as the equivalent of the Lamb and the AC-Stark shifts. 

The presence of a nonlinear Kerr effect of the same order as the usual first-order frequency shifts paves the way to interesting future studies. For instance, two-photon interactions could be used to improve qubit readout protocols in the dispersive regime. Given that the Kerr effect provides an intracavity squeezing term, external sources of squeezed radiation would be no longer necessary~\cite{Barza2014}. The nonlinear Kerr effect could also be applied in producing nonclassical states (e.g., squeezed cat states) and in the enhancement of the multiphoton blockade~\cite{Mira2013}.

\section{Approaching the spectral collapse}
\label{sec_collapse}
In this section, we discuss the minimal value of the qubit-oscillator interaction strength required to approach the spectral collapse of the two-photon QRM. Then, we consider higher-order terms derived in Sec~\ref{sec_scheme} in order to understand the physical response of the system when the spectrum of the two-photon QRM collapses.

So far, the spectral collapse of the two-photon QRM has been mostly studied considering the interaction term $\sigma_x ( a^2 + {a^\dagger}^2 )$~\cite{Travenec2012, Albert2011,Chen2012,Duan16,Twophot_ions,Puebla17}. In that case, the spectral collapse takes place for a critical value of the coupling strength $g_{\text{col}}=\omega_c/2$. The circuit design proposed in Sec.~\ref{sec_scheme} is instead described by a full quadratic coupling $\sigma_x ( a + a^\dagger)^2$, similar to a recent proposal to implement generalized Rabi models in cold atoms~\cite{Schneeweiss17}. The full quadratic coupling is the actual physical representation of the qubit-photon interaction, as the qubit is coupled to the square of the electric or magnetic field of the cavity mode. When the full quadratic coupling is taken into account, the critical value of the coupling strength is given by $g_{\text{col}}=\omega_c/4$, which is a more attainable quantity. 

\begin{figure}[]
\centering
\includegraphics[angle=0, width=0.4\textwidth]{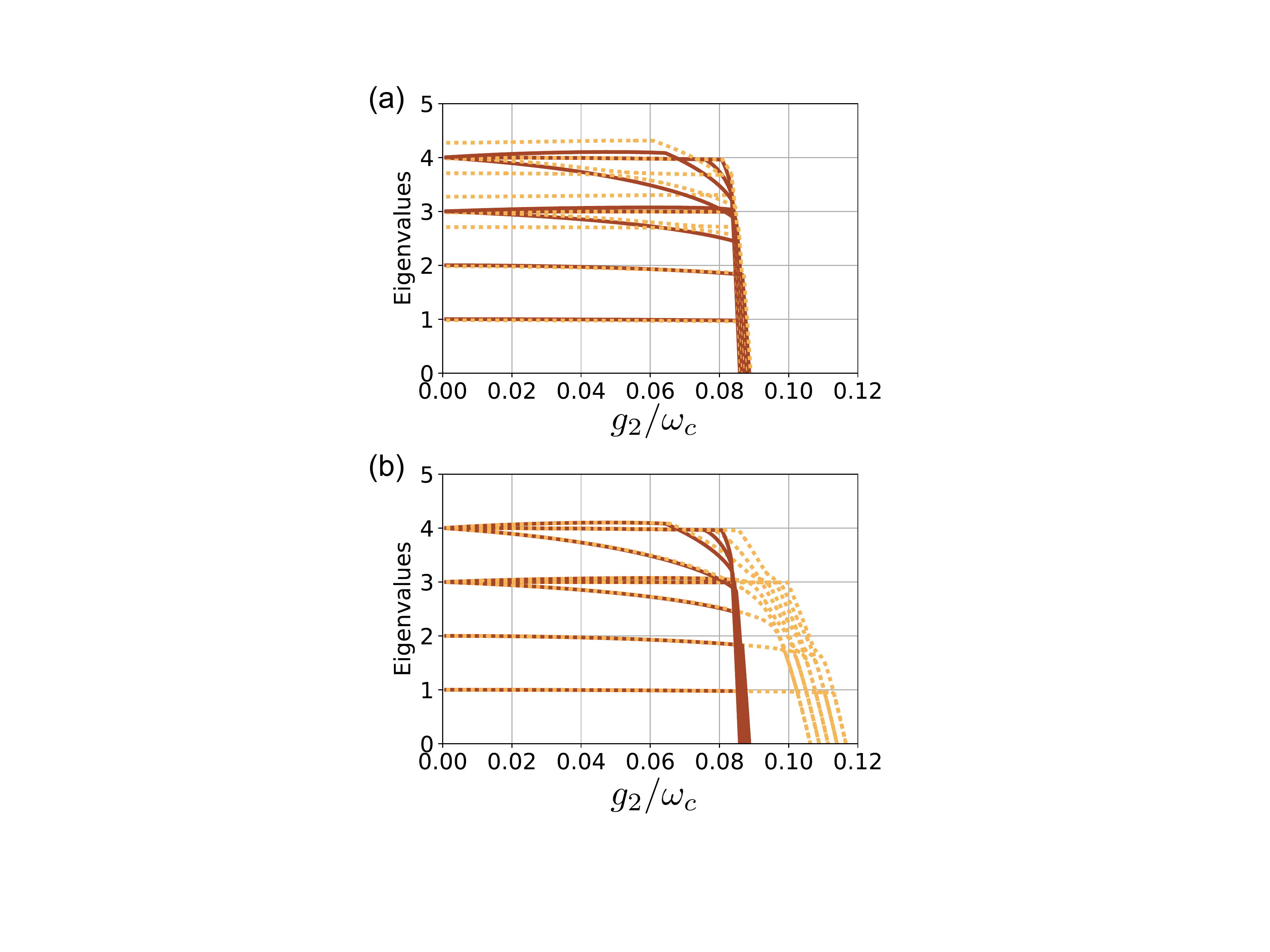}
\caption{\label{Dicke_coll} Numerical solutions for the eigenenergies of two-photon interaction models, as function of the coupling strength, in the three-qubit case [$N=3$ in Eq.~\eqref{multiqubit}]. We consider resonant interactions $\omega_q = 2\omega_c$. For the sake of clarity, only the first $10$ energy levels are displayed. A cut-off $n=150$ for the maximum allowed number of photons has been imposed. (a) Eigenenergies of the two-photon QRM (solid red line) and of the model obtained adding inter-spin couplings of strength $J=0.2\omega_c$. The presence of inter-spin interactions modifies the spectrum in the SC regime but becomes irrelevant close to the collapse point. (b) Energy levels of the two-photon QRM (solid red line) and of the model obtained by adding the quartic coupling term with $g_4/g_2 = 10^{-3}$. The quartic term is negligible in the SC regime, while around the collapse point it suddenly becomes necessary to correctly assess the system ground state.}
\end{figure}

In the following, we provide an intuitive explanation for the appearance of the spectral collapse. Furthermore, we show that the addition of more qubits reduces the critical value of the individual qubit coupling strength needed to reach the collapse point. When multiple qubits are involved, the system Hamiltonian can be written as
 \begin{equation}
 \mathcal{H} = \omega_c a^\dagger a + g_2 N S_x \left(a^\dagger + a \right)^2 + \mathcal{H}_{\text{spin}},
 \label{multiqubit}
 \end{equation}
 in which $N$ is the number of qubits and $S_x=\frac{1}{N}\sum_{i=1}^N \sigma^i_x $ is the normalized collective spin operator. In the last term $H_{\text{spin}}$, we gathered all terms including only spin operators $H_{\text{spin}}=\frac{\omega_q}{2}\sum_{i=1}^N \sigma^i_x$, where the index $i$ identifies each qubit.
Let us rewrite the system Hamiltonian in terms of the field quadratures,
$x = \frac{1}{\sqrt{2\omega}}\left( a^\dagger + a \right)$ and $p = i\sqrt{\frac{\omega}{2}}\left( a^\dagger - a \right)$,
\begin{equation}
\mathcal{H} = \frac{\omega^2}{2}\left[ 1 + \frac{4g_2 N}{\omega_c} S_x \right]x^2 + \frac{1}{2}p^2 + \mathcal{H}_{\text{spin}}.
\label{H_quadratures}
\end{equation}
The norm of $S_x$ is bound to be between $-1$ and $1$, so when $g_2 \geq g_{\text{col}}= \frac{\omega_c}{4N} $ there exist spin states for which the optical field potential in Eq.~\eqref{H_quadratures} is flat or even unbounded from below. Being $\mathcal{H}_{\text{spin}}$ a finite-norm operator, it cannot compensate this divergence and so the spectral collapse takes place. The dependence $1/N$ of the collapse value $g_\text{col}$ relaxes the experimental constraints on the single qubit coupling strength, as including more qubits in the circuit design could result less challenging than increasing the individual coupling parameters.

Notice that the previous argument is still valid for more structured spin Hamiltonians. It has been recently shown that spin-spin interactions inhibit the onset of the phase transition in the thermodynamic limit of the Dicke model~\cite{Jaako16}. This is not the case for the spectral collapse of the two-photon Dicke model, which takes place for any value of $N$. In Fig.~\ref{Dicke_coll}(a) we show the system spectrum for $N=3$, where we compare the Dicke model of Eq.~\eqref{H_quadratures} (solid red line) to the model obtained adding to $\mathcal{H}_\text{spin}$ interspin interactions of the kind $J \sum_{i=1}^{N-1} \sigma_x^i\sigma_x^{i+1}$ (dashed yellow line).
The system spectrum collapses for values of the individual qubit-oscillator coupling smaller than $0.09 \omega_c$, well within the reach of present superconducting-circuit technology. The interspin coupling modifies the spectrum for low coupling strengths, but it does not qualitatively modify the system spectrum in proximity of the spectral collapse.

Having theoretically shown that the collapse point can be achieved in practical implementations, let us analyze the spectrum of the circuit scheme proposed in Sec~\ref{sec_scheme} at and beyond the collapse point. In order to show that the two-photon QRM Hamiltonian of Eq.~\eqref{original2phHam} effectively describes the full circuit model Eq.~\eqref{eq:H_sq}, we expanded the cosinusoidal potential of the SQUID and kept terms up to quadratic order in the SQUID internal phase. When no current biases the SQUID, odd-power contributions vanish, so that the first correction is given by the quartic term $U_\text{4P} = g_4 \sigma_x \left( a + a^\dagger \right)^4$. The prefactor of the quartic term is much smaller than the two-photon coupling strength ($g_4/g_2 \sim 10^{-3}$), hence in the SC regime it can be safely neglected. However, when the two-photon QRM undergoes a spectral collapse the spectrum becomes unbounded from below and higher order terms must be taken into account in order to obtain a physical model of the system. In Fig~\ref{Dicke_coll}(b) we compare the eigenenergies of the two-photon QRM (solid red line) with the model obtained by adding the quartic term  (dashed yellow line). The two spectra coincide up to the spectral collapse ($g_2 = \omega_c/12$). Beyond the collapse point, the quartic term suddenly starts to play a relevant role, providing the model with a physical ground state. If the coupling strength keeps increasing, the spectrum also collapses in presence of the quartic coupling term, implying that higher orders in the expansion will become increasingly relevant.

\section{Discussion and perspectives}
\label{sec_conclusions}
Let us now briefly summarize our results before commenting on future perspectives.  We proposed a circuit scheme where dipolar and nondipolar interaction terms can be selectively activated. The coupling strength of one- and two-photon interactions can be independently tuned in real-time by adding static current and flux biases to the main SQUID loop, allowing the exploration of both strong and ultrastrong couplings of nondipolar interactions.  In particular, we focused on the two-photon quantum Rabi model, which can be implemented with a full quadratic coupling.

We presented the first input-output analysis of two-photon interactions in the SC regime, highlighting fundamental differences with the Jaynes-Cummings physics. For a weak coherent cavity driving, the two-photon Rabi splitting is not visible, as the first-excited level of the model is a purely photonic state, unlike the polaritonic splitting for dipolar interactions. In this regime, the system represents a single-photon source that is robust toward qubit decoherence. For strong cavity driving, the two-photon QRM saturates but the presence of the qubit is still relevant in the transmission spectrum, which is characterized by a double-peak structure. We provided an analytical description for this behavior, which bears the signature of level quantization also for intense fields. When the external driving is applied to the qubit, the transmission spectrum is characterized by different selection rules, established by the conservation of the parity of the photon number. The output photon statistics shows a pronounced two-photon blockade effect which takes place as a first-order phenomenon, which is then visible also for few-photon drivings.

We analytically derived effective Hamiltonians that include perturbative corrections when the rotating-wave approximation is broken. In particular, we considered the Bloch-Siegert and the dispersive regimes. The perturbative corrections are given by Lamb, AC-Stark and qubit-state-dependent Kerr shifts. The latter represents a fundamental difference with respect to dipolar interactions that could lead to practical advantages in quantum information tasks.

Finally, we analyzed the system spectrum at the collapse point of the two-photon QRM. We show that by adding more qubits to the circuit the spectral collapse can be reached with relatively small coupling strength. When the spectral collapse takes place, higher-order interaction terms must be included in order to avoid divergences of the eigenenergies, suggesting that they could become relevant to the system dynamics.

Our work paves the way to the experimental and theoretical exploration of nondipolar interactions in microwave quantum photonics, both in the strong and in the ultrastrong coupling regime. Immediate follow-up studies will focus on the study of the modification introduced by two-photon interactions on standard cavity QED phenomena like superradiance, which is the radiation rate enhancement experienced by multiple qubits coupled with the same optical mode. In terms of applications, our work could be relevant for the generation of squeezed cat states, namely, entanglement generation between qubits and highly-populated cavity squeezed states. Furthermore, these results could be generalized to obtain nondipolar interactions in waveguide-QED, where atoms interact with a continuum of modes, resulting in nondipolar spin-boson models. Finally, nondipolar couplings represent a novel kind of interactions that could be exploited in quantum simulation protocols or in the implementation of many-body phenomena with quantum fluids of light~\cite{Carusotto13}.

\begin{acknowledgments}
S.F.~acknowledges support from the French Agence Nationale de la Recherche (SemiQuantRoom project, No.~ANR-14-CE26-0029) and from the PRESTIGE program, under the Marie Curie Actions-COFUND of the FP7. D.Z.R.~acknowledges support from S\~{a}o Paulo Research Foundation (FAPESP) Grants No.~2013/23512-7 and No.2014/24576-1. E.S. and E.R. acknowledge support from Spanish MINECO/FEDER FIS2015-69983-P, Basque Government IT986-16 and UPV/EHU grant EHUA15/17. P.~F.-D. is supported by the Beatriu de Pin\'os fellowship.

\end{acknowledgments}

\appendix
\section{Input-output theory}
\label{inout}
We assume that the cavity is coupled to one-dimensional waveguides via the standard interaction 
$ V_a = \sqrt{\frac{\gamma}{2\pi}}\int d\omega \left[  b_i(\omega)  a ^\dagger +  b_i^\dagger(\omega)  a \right]$, where $\gamma$ is the cavity dissipation rate and $ b_i(\omega)$ are annihilation operators for the input-output fields~\cite{Walls94}. In the same way, the interaction Hamiltonian of the qubit with the input waveguide is given by
$ V_\sigma = \sqrt{\frac{\gamma_q}{2\pi}}\int d\omega \left[  c(\omega)  \sigma_+ ^\dagger +  c^\dagger(\omega)  \sigma_- \right]$, where $ c(\omega)$  are annihilation operators of the input fields, and they also model qubit dissipative decay.
These interactions lead to the master equation
\begin{equation}
\dot{\rho}(t) = i\left[ \rho(t), \mathcal{H} \right] + \mathcal{L}_a \rho(t) + \mathcal{L}_\downarrow \rho(t) + \mathcal{L}_\phi \rho(t),
\end{equation}
where $\mathcal{H}= \mathcal{H}_S + \mathcal{H}_d $ includes the system and driving Hamiltonians. In the SC regime, the system Hamiltonian is given by Eq.~\eqref{2phSC} for the two-photon case, and by the JC model in the one-photon case. We consider $\mathcal{H}_d = 2 D \cos(\omega_d t) \left( s^\dagger + s \right)$, with $s= a$ ($s=\sigma_-$) for cavity (qubit) driving. The effective intensity of the driving $D= \beta \sqrt{\gamma}$ is given by decay rates and by the amplitude $\beta$ of the input coherent field. Notice that $\beta$ is normalized in such a way that $n_{\text{in}}=|\beta|^2$ represents the flux of photons per second of the input field. The cavity decay due to the interaction with an external waveguide is described by the Louivillian operator $\mathcal{L}_a \rho(t) = \gamma \mathcal{D}[a] \rho(t)$, where 
$\mathcal{D}[\mathcal{O}]\rho = \mathcal{O}\rho \mathcal{O}^\dagger - \frac{1}{2} \left\{ \rho, \mathcal{O}^\dagger \mathcal{O}\right\}$. The model also includes qubit decay 
$\mathcal{L}_\downarrow \rho(t) = \gamma \mathcal{D}[\sigma_-] \rho(t)$ and pure dephasing $\mathcal{L}_\phi \rho(t) = \gamma_\phi \mathcal{D}[\sigma_z] \rho(t)$. The output fields are obtained by numerically solving the master equation and by using the input-output relation $a_{\text{out}} (t) = a_{\text{in}} (t) + \sqrt{\gamma} a(t)$ ~\cite{Walls94}.

\section{Strong-driving limit}
\label{StongDrive}
In the SC regime, for resonant interaction ($\omega_q=2\omega$), the two-photon QRM Hamiltonian can be written in the interaction picture as
$\mathcal{H}_I = g_{2}(  \sigma_+ a^2 + \sigma_- {a^\dagger}^2 )$. This Hamiltonian preserves a continuous symmetry given by a weighted excitation-number operator $C_{\text{2ph}} = \sigma_z + 2a^\dagger a$. Notice that when the USC regime is reached, this continuous symmetry breaks down into a discrete one $\Pi_{\text{2ph}} = \exp\{ i\pi a^\dagger a \}$, which corresponds to the parity of the photon number. However, in the SC regime the continuous symmetry allows to diagonalize the Hamiltonian. Beyond the ground $(\ket{\text{g},0})$ and first-excited  $(\ket{\text{g},1})$ states, the system eigenstates are given by the doublets 
$\ket{\psi_n^\pm}=\left( \ket{\text{g}, n+2} \pm \ket{\text{e},n} \right)/\sqrt{2}$. Setting the ground-state energy to zero, the energy eigenvalues corresponding to the doublets are given by 
$E_n^\pm= \omega_c (n+2) \pm g_{2}\sqrt{(n+1)(n+2)}$.

In the case of cavity driving, the driving Hamiltonian is given by $\mathcal{H}_d = D( e^{i\omega_d t} a^\dagger + e^{-i\omega_d t} a)$. Let us rewrite the ladder operators in the basis of the two-photon QRM doublets 
\begin{equation}
\langle \psi_n^s | a | \psi_{n^\prime}^{s^\prime}\rangle = \frac{1}{2}\left[ \sqrt{n^\prime+2} + s s^\prime \sqrt{n^\prime} \right] \delta(n+1, n^\prime),
\end{equation}
where $s=\pm 1$.
In the limit of large $n$, we can approximate
\begin{equation}
\langle \psi_n^s | a | \psi_{n^\prime}^{s^\prime}\rangle \approx  \sqrt{n^\prime} \delta(n+1, n^\prime) \delta(s, s^\prime).
\end{equation}
It follows that for highly populated states the cavity driving induces transitions between the eigenstates of neighbouring doublets, without mixing the lower with higher states of each doublet. Furthermore, the $\sqrt{n^\prime}$ factor shows that the energy ladder becomes harmonic. Indeed, we can calculate the energy difference $\Delta_n^\pm= E_n^\pm - E_{n-1}^\pm$ between lower or higher eigenstates of neighbouring doublets 
\begin{equation}
 \Delta_n^\pm = \omega_c \pm g_2 \left[ \sqrt{(n+1)(n+2)} -  \sqrt{n(n+1)} \right].
\end{equation}
For highly excited states we obtain   $\lim_{n\rightarrow \inf}  \Delta_n^\pm  =\omega \pm g$, which explains the double-peak structure in the strong-driving limit of Fig.~\ref{cavityDrive}. 
A similar analysis holds for the JC model, with the difference that for one-photon coupling the energy splitting between successive lower or higher eigenstates is degenerate $ \Delta_n^\pm=\omega$, and so the spectrum for highly excited states is well approximated by an empty cavity.

\end{document}